\documentclass[twocolumn,showpacs,preprintnumbers,amsmath,amssymb,prl]{revtex4}

\usepackage{graphicx,mathrsfs,amssymb}% Include figure files
\usepackage{dcolumn}% Align table columns on decimal point
\usepackage{bm}% bold math

\begin{document}

\newcommand{\trace}[1]{\ensuremath{\mathrm{Tr}\{ #1 \}}}
\newcommand{\bra}[1]{\ensuremath{\left< #1 \right|}}
\newcommand{\ket}[1]{\ensuremath{\left| #1 \right>}}
\newcommand{\mean}[1]{\ensuremath{\left< #1 \right>}}

\newcommand{\ud}{\mathrm{d}}
\newcommand{\E}{\mathscr{E}}
\newcommand{\kbf}{\mathbf{k}}
\newcommand{\rbf}{\boldsymbol{r}}
\newcommand{\kappabf}{\boldsymbol{\kappa}}
\newcommand{\Ebf}{\boldsymbol{E}}
\newcommand{\mbf}[1]{\boldsymbol{#1}}
\newcommand{\aop}{\hat{a}}

\title{Single-photon excitation of a coherent state: catching the elementary step of stimulated light
emission}

\author{Alessandro Zavatta}
\author{Silvia Viciani}
\author{Marco Bellini}

\email{bellini@inoa.it}

\affiliation{Istituto Nazionale di Ottica Applicata,\\L.go E. Fermi, 6, I-50125,
Florence, Italy \\LENS and Department of Physics, University of Florence, I-50019 Sesto
Fiorentino, Florence, Italy}

\date{\today}

\begin{abstract}
When a single quantum of electromagnetic field excitation is added to the same
spatio-temporal mode of a coherent state, a new field state is generated that exhibits
intermediate properties between those of the two parents. Such a single-photon-added
coherent state is obtained by the action of the photon creation operator on a coherent
state and can thus be regarded as the result of the most elementary excitation process
of a classical light field. Here we present and describe in depth the experimental
realization of such states and their complete analysis by means of a novel ultrafast,
time-domain, quantum homodyne tomography technique clearly revealing their
non-classical character.
\end{abstract}


\maketitle

\section{Introduction}

A coherent state $\ket{\alpha}$, the eigenstate of the photon annihilation operator
$\aop\ket{\alpha}=\alpha\ket{\alpha}$, is the closest analogue to a classical light field and
exhibits a Poisson photon number distribution with an average photon number of $|\alpha|^2$
and a width of $|\alpha|$. Coherent states posses well defined amplitude and phase, whose
uncertainties are the minimum permitted by the Heisenberg uncertainty principle. On the
contrary, a Fock state, the eigenstate of the photon number operator
$\aop^\dag\aop\ket{n}=n\ket{n}$, contains a perfectly defined number of quanta of field
excitation and is strictly quantum-mechanical, with no classical analog. Moreover, being the
intensity of a Fock state defined without uncertainty, its phase is completely undefined.

Single-photon Fock states have been recently generated by means of conditional preparation
techniques and homodyne tomography has been used to completely characterize them with the
reconstruction of an associated Wigner function clearly exhibiting negative
values~\cite{lvovsky01,pra04}. Displaced Fock states, obtained by mixing a coherent state with
a single photon upon a highly reflecting beam splitter as in Ref.~\cite{lvovsky02:pra}, have
also been investigated with tomographic techniques which have shown the non-Gaussian character
of their marginal distributions and negative values of the Wigner function. Other
non-classical states have been recently produced starting from a squeezed vacuum and with the
controlled subtraction of a single photon: in this case the marginal distributions clearly
showed a squeezed and non-Gaussian character but the preparation and detection efficiency was
not high enough to reconstruct a negative-valued Wigner function~\cite{wenger04}.

We have recently reported~\cite{science04} the experimental generation of a new kind of
non-classical field states and their tomographic analysis based on time-resolved
homodyne detection, which has allowed us to observe both the squeezed character and the
negativity of the associated Wigner function. These so-called single-photon-added
coherent states are produced whenever a single photon is injected in the same
spatio-temporal mode of a coherent state and are shown to exhibit a mix of the
characteristics of both parents. In particular, by simply varying the contribution of
the initial coherent state, the character of the final state can be continuously tuned
between completely quantum and almost completely classical.

Here we report an in-depth tomographic analysis of such states based on new and more
accurate experimental data and comprising a complete comparison of the reconstructed
density matrix elements with those expected from a fully developed theory. This has
allowed us to follow in a much more detailed way the evolution of the generated state
from the particle-like one, characterized by a circularly symmetric and negative-valued
single-photon Wigner function, through a squeezed intermediate region characterized by
the gradual appearance of a phase, towards the wave-like classical coherent state.

In addition to the interesting physical properties of single-photon-added coherent
states, the ability to generate, manipulate and characterize such states can be useful
for possible future applications in the engineering of quantum
states~\cite{lund04,dakna98} and in quantum information protocols~\cite{wenger04}.

\section{Properties of the SPACS}

In 1991, Agarwal and Tara~\cite{agarwal91} introduced a new class of states,defined by the
repeated ($m$ times) application of the photon creation operator to the coherent state,
\begin{equation}
\ket{\alpha,m}=k_{\alpha,m}\,\aop^{\dag m}\ket{\alpha}, \label{eq:pacs}
\end{equation}
with $k_{\alpha,m}=[m!\,L_m(-|\alpha|^2)]^{-1/2}$ a normalization factor where $L_m(x)$
is the $m$th-order Laguerre polynomial and $m$ is an integer. Such photon-added
coherent states (PACSs) essentially represent the result of successive elementary
one-photon excitations of a classical coherent field and occupy an intermediate
position between the Fock and the coherent states, reducing to the two limit cases for
$\alpha\rightarrow0$ or $m\rightarrow0$, respectively. From the expansion of PACSs in
terms of Fock states, it can be easily seen that they essentially correspond to a
shifted version of a coherent state where all the $\ket{n}$ terms with $n<m$ are
missing and that all the elements of the corresponding density matrix are re-scaled and
displaced towards higher indices $\rho_{i,j}\rightarrow\rho_{i+m,j+m}$, leaving all the
elements with $i,j<m$ void.

\begin{figure}[h]
\includegraphics[width=85mm]{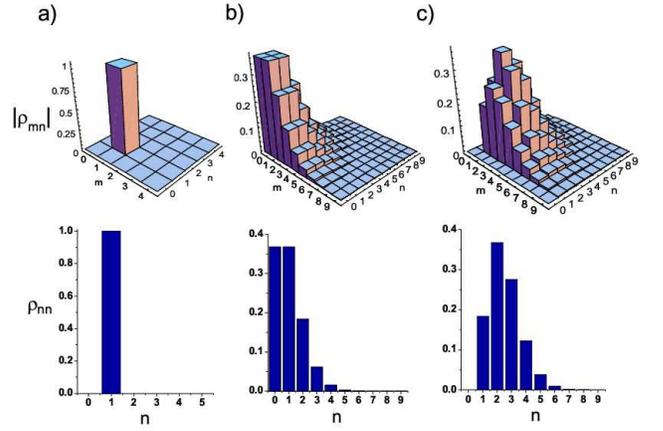}
\caption{(color online) Theoretical density matrix elements and photon number distributions
for: a) the single-photon Fock state $\ket{1}$, b) the coherent state $\ket{\alpha}$ (with
$|\alpha|=1$) and c) the corresponding SPACS $\ket{\alpha,1}$ obtained by emission of a single
photon in the mode of the coherent state.\label{fig:matrix_theo}}
\end{figure}

When just a single quantum of field excitation is added to a coherent field, the
single-photon-added coherent state (SPACS) reads as:
\begin{equation}
\ket{\alpha,1}=\frac{\aop^{\dag}\ket{\alpha}}{\sqrt{1+|\alpha|^2}}
\label{eq:spacs}
\end{equation}
and can be also viewed as the superposition of a displaced single-photon Fock state and
a coherent state~\cite{agarwal91}. SPACSs can be expanded in terms of Fock states as:
\begin{equation}
\ket{\alpha,1}=\frac{e^{-\frac{|\alpha|^2}{2}}}{\sqrt{1+|\alpha|^2}}\sum_{n=0}^{\infty}\frac{\alpha^n}{\sqrt{n!}}\sqrt{n+1}\ket{n+1}
\label{eq:spacs_n}
\end{equation}
where the lack of the vacuum term contribution is evident. Accordingly, the density
matrix elements for the SPACSs are:
\begin{equation}
\rho_{i,j}^{\ket{\alpha,1}}=\frac{i
j}{\sqrt{i!j!}}\frac{e^{-|\alpha|^2}}{1+|\alpha|^2}\alpha^{(i-1)} \alpha^{*(j-1)}
\label{eq:spacs_rho}
\end{equation}
and the effect of single-photon excitation can be readily observed in the plots of
Fig.~\ref{fig:matrix_theo} where the absolute value of the theoretical matrix elements and the
photon number distributions (their diagonal elements) are reported for the single-photon Fock
state, for a coherent state with $\alpha=1$ and for the corresponding SPACS.

Unlike the operation of photon annihilation, which maps a coherent state into another coherent
state ($\aop \ket{\alpha}=\alpha \ket{\alpha}$), i.e. a classical field into another classical
field, the single-photon excitation of a coherent state changes it into something quite
different, especially for low values of $\alpha$, where the absence of the vacuum term has a
stronger impact. In the extreme case of an initial vacuum state $\ket{0}$, the addition of one
photon indeed transforms it into the very non-classical single-photon Fock state $\ket{1}$,
which exhibits negative values of the Wigner function around the origin. More generally, the
Wigner function for a single-photon-added coherent state of arbitrary amplitude $\alpha$ can
be expressed as:
\begin{equation}
W(z)=\frac{-2(1-|2z-\alpha|^2)}{\pi (1+|\alpha|^2)}e^{-2|z-\alpha|^2}
\label{eq:spacs_wig_teo}
\end{equation}
(where $z=x+iy$) and can clearly become negative, a proof of its non-classical
character, whenever the condition
\begin{equation}
|2z-\alpha|^2<1
\end{equation}
is satisfied. Thus, in general, the application of the creation operator $\aop^\dag$,
changes a completely classical coherent state into a quantum state with a varying
degree of non-classicality, which becomes more evident the smaller the initial
amplitude of the $\ket{\alpha}$ state. If the amplitude $\alpha$ is gradually increased
from zero, the smooth transition from an initial purely quantum state (the
single-photon Fock state) towards a classical coherent one (with the birth and the
gradual appearance of a well defined phase) can be achieved.
\begin{figure}[h]
\includegraphics[width=85mm]{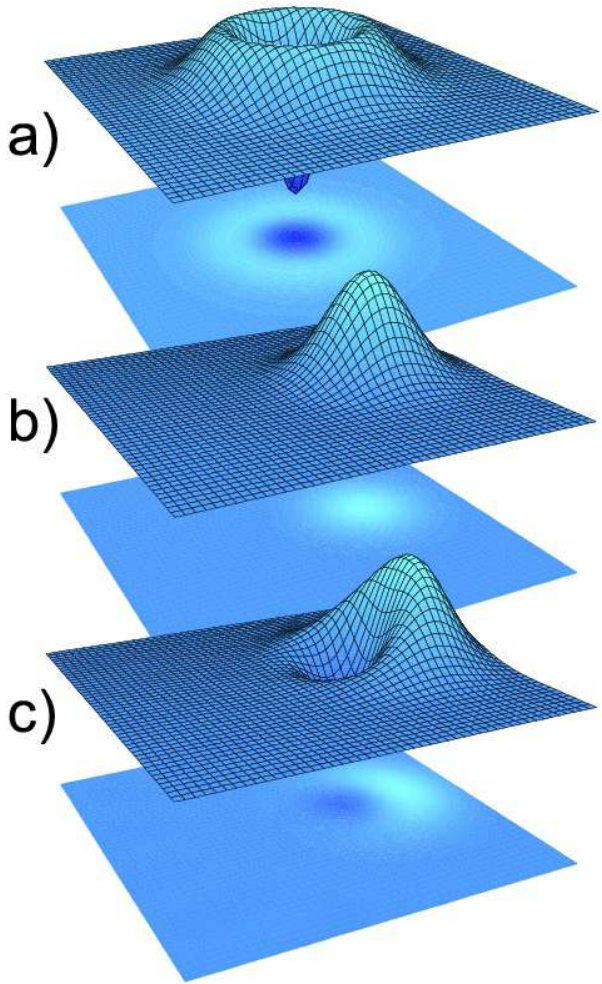}
\caption{(color online) Theoretical Wigner function for: a) the single-photon Fock state
$\ket{1}$, b) the coherent state $\ket{\alpha}$, c) the SPACS $\ket{\alpha,1}$. A value of
$|\alpha|^2=1$ is used. \label{fig:wigner_theo}}
\end{figure}
In addition to the negativity of the Wigner function, SPACSs also exhibit a definite squeezing
in their field quadratures that can be readily observed. Given a field quadrature
$\hat{x}_\theta=\frac{1}{2}(\aop e^{-i\theta}+\aop^\dag e^{i\theta})$, its mean value is:
\begin{equation}
\mean{x_\theta}_\alpha= \bra{\alpha,1}\hat x_\theta\ket{\alpha,1}=
\frac{|\alpha|(2+|\alpha|^2)\cos(\theta)}{1+|\alpha|^2}
\label{eq:spacs_mean_teo}
\end{equation}
and its fluctuations amount to:
\begin{equation}
[\Delta x_\theta]^2_\alpha= \mean{x^2_\theta}_\alpha-\mean{x_\theta}^2_\alpha=
%\frac{2+(1+|\alpha|^2)^2-2|\alpha|^2\cos{2\theta}}{4(1+|\alpha|^2)^2}
\frac{1}{4}+\frac{1-|\alpha|^2\cos{(2\theta)}}{2(1+|\alpha|^2)^2}
\label{eq:spacs_var_teo}
\end{equation}
Clearly, the quadrature obtained by choosing $\theta=0$ exhibits reduced fluctuations with
respect to the coherent state for $|\alpha|>1$, and is thus squeezed. It is interesting to
note that, differently from Fock and Gaussian squeezed states, SPACSs combine both the key
features normally associated to quantum states: the negativity of the Wigner function and the
reduced fluctuations along one quadrature.

\section{Generation of the SPACS}

SPACSs can be generated by injecting a coherent state $\ket{\alpha}$ into the signal
mode of an optical parametric amplifier and exploiting the stimulated emission of a
single down-converted photon into the same mode. Differently from conventional optical
amplification where a coherent state is converted to another coherent state, here a
low-gain regime of the amplifier and a conditioning of the state based on measurements
on the idler mode are required to exactly select the one-photon excitation term (i.e.
to avoid higher-order excitations which cannot be discriminated by our single-photon
detectors and to exclude the vacuum contribution). Hence, in order to make sure that
single-photon emission has taken place in the signal channel, one can use a conditional
preparation technique which guarantees the generation of the target state every time
that a single photon is detected in the correlated idler mode.

The Hamiltonian for the parametric amplifier reads as:
\begin{equation}
H=i\hbar\chi (\aop_s^\dag \aop_i^\dag -\aop_s \aop_i)
\end{equation}
where $\aop_i$ ($\aop_s$) is the annihilation operator for the idler (signal) mode and $\chi$
is proportional to the amplitude of the (classical) pump and to the second-order
susceptibility of the medium. The time evolution of an initial state $\ket{\psi(0)}$ is thus
described by:
\begin{eqnarray}
\ket{\psi(t)}=e^{-i \frac{Ht}{\hbar}}\ket{\psi(0)}=\nonumber \\=e^{\chi t (\aop_s^\dag
\aop_i^\dag -\aop_s \aop_i)}\ket{\psi(0)}.
\end{eqnarray}

If the parametric gain is kept sufficiently low ($g=\chi t\ll 1$), which is always the case in
our experimental situation, the final output state can be approximated as
\begin{equation}
\ket{\psi(t)} = [1+g (\aop^\dag_s \aop^\dag_i- \aop_s \aop_i)]\ket{\psi(0)}.
\label{eq:pacs_gen}
\end{equation}
By letting a seed coherent field $\ket{\alpha}_s$ enter the parametric crystal in the signal
mode, while vacuum ($\ket{0}_i$) enters in the idler channel, the final state becomes
\begin{eqnarray}
\ket{\psi} = [1+g (\aop^\dag_s \aop^\dag_i- \aop_s \aop_i)]\ket{\alpha}_s\ket{0}_{i}=\nonumber
\\ =\ket{\alpha}_s\ket{0}_{i}+g \aop^\dag_s\ket{\alpha}_s\ket{1}_{i}
\label{eq:pacs_cond}
\end{eqnarray}
and the output signal mode will mostly contain the original coherent state, except for the few
cases when the state $\ket{1}_{i}$ is detected in the idler output mode. These relatively rare
detection events, which take place with a probability proportional to $|g|^2(1+|\alpha|^2)$,
project the signal state onto the desired SPACS $\ket{\alpha,1}_s$, corresponding to the
stimulated emission of one photon in the same mode of $\ket{\alpha}$. Note that when the input
state is of the form $\ket{0}_{s}\ket{0}_{i}$, i.e. no seed coherent field is injected into
the crystal, spontaneous parametric down-conversion takes place starting from the input vacuum
fields, and pairs of entangled signal and idler photons with random (but mutually correlated)
phases are produced in the crystal in the state $\ket{1}_{s}\ket{1}_{i}$ with a low
probability proportional to $|g|^2$. In this case, the detection of a single photon in the
idler mode projects the signal state onto a single-photon Fock state and, by following the
evolution of the final quantum state while the amplitude $\alpha$ increases from zero, one can
witness the gradual transition from the spontaneous to the stimulated regimes of light
emission with the smooth transformation of a single photon (particle-like) state towards a
coherent (wave-like) one.

Quite interestingly, one can obtain an absolute calibration of the amplitude of the
seed coherent field $\ket{\alpha}_s$ injected in the SPDC signal mode by measuring the
rate of counts in the idler channel and comparing them to the un-seeded case. As stated
above, the ratio of such rates equals $(1+|\alpha|^2)$ and this is clearly due to the
enhancement of emission probability characteristic of stimulated emission in bosonic
fields. The same scheme was originally proposed by Klyshko~\cite{klyshko77} as a
metrological tool for absolute radiance measurements~\cite{kitaeva79,migdall99}.

For low $\alpha$ values, one can truncate the above expressions to the first two terms
of the coherent state expansion in the number state basis. In this case the final state
becomes:
\begin{eqnarray}
\ket{\psi}\approx[1+g (\aop^\dag_s \aop^\dag_i- \aop_s \aop_i)]\, (\ket{0}_{s}+\alpha
\ket{1}_s)\, \ket{0}_i=\nonumber
\\=(\ket{0}_s+\alpha \ket{1}_s)\,\ket{0}_i+g (\ket{1}_s+\sqrt{2}\alpha
\ket{2}_s)\,\ket{1}_i
\label{eq:pacs_cond_trunc}
\end{eqnarray}
and the conditioning thus reduces the signal state to the coherent superposition
$\ket{1}_s+\sqrt{2}\alpha \ket{2}_s$. Note that such a coherent superposition of two number
states possesses a well defined phase and is highly non-classical, completely missing the
contribution of the vacuum.

The same state as the one described by (\ref{eq:pacs_cond}) and (\ref{eq:pacs_cond_trunc}) has
been recently generated and used by Resch et al.~\cite{resch02} to generate an arbitrary
superposition of zero- and one-photon states. In that case, however, the conditioning was
performed upon the detection of a single photon in the same mode $\ket{1}_s$ of the input
coherent state, hence the final state was completely different from the ones investigated
here, and of the form $(\alpha\ket{0}_i+g\ket{1}_i)$. The injection of a single photon instead
of a coherent state as a seed for conditional parametric amplification has also been
investigated in \cite{ou90:josab} and experimentally demonstrated \cite{demartini00}, with the
amplification to a $\ket{2}_s$ Fock state in the context of quantum cloning.

\section{Experimental setup}

\begin{figure}[h]
\includegraphics[width=85mm]{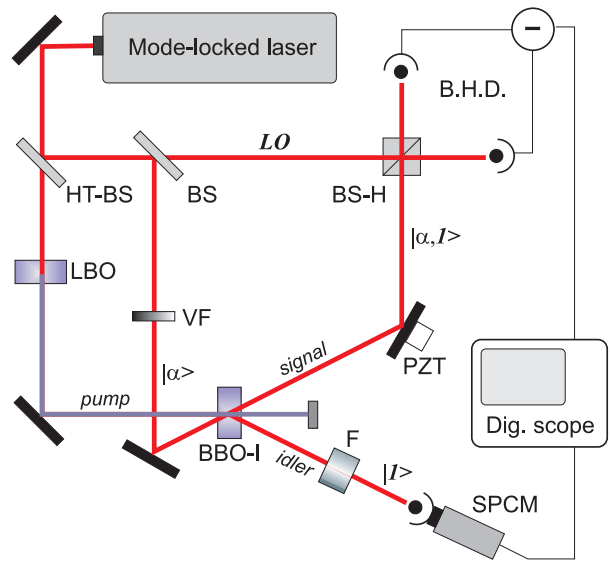}
\caption{(color online) Experimental apparatus: HT-BS high transmission beam-splitter,
LBO lithium triborate crystal, BS and BS-H 50\% beam-splitters, VF variable attenuation
filter, BBO-I type-I $\beta$-barium borate down-converter crystal, PZT piezoelectric
transducer, B.H.D. balanced homodyne detector, F spectral and spatial filters, SPCM
single photon counting module, LO local oscillator. \label{fig:setup}}
\end{figure}

The experimental apparatus used to generate and analyze the SPACS is schematically
drawn in Fig.\ref{fig:setup}. A mode-locked Ti:sapphire laser, emitting 1-2 ps long
pulses at 786 nm and at a repetition rate of 82 MHz is used as the primary source. The
laser pulses are frequency doubled to 393 nm in a 13-mm long LBO crystal which thus
produces the pump pulses for parametric down-conversion in a 3-mm thick, type-I BBO
crystal. The crystal is slightly tilted from the collinear configuration in order to
obtain an exit cone beam with an angle of $\sim 3^\circ$ from which symmetric signal
and idler modes are roughly selected by means of irises placed at about 70 cm from the
crystal.

In order to non-locally select a pure state on the signal channel, idler photons undergo
narrow spatial and frequency filtering before detection; indeed the nonlocally-prepared signal
state will only approach a pure state if the filter transmission function is much narrower
than the momentum and spectral widths of the pump beam generating the SPDC
pair~\cite{ou97,lvovsky02:epjd,bellini03,viciani04}. The idler beam is thus passed through a
pair of etalon interference filters which perform a narrow (50 GHz) spectral selection and is
then coupled into a single-spatial-mode fiber before impinging onto a single photon counting
module (Perkin-Elmer SPCM AQR-14).

The weak coherent state $\alpha$ is obtained by controlled attenuation (VF in the
figure, composed of a polarizer and a half-wave plate) of a small portion of the laser
emission which is fed into the signal mode of the parametric crystal and is then
directed to a 50\% beam-splitter (BS-H in figure). Here it is overlapped with a second
(intense) coherent state (again obtained from a portion of the original laser pulses)
which is spatially mode-matched to the conditionally-prepared SPACS by the insertion of
appropriate lens combinations (not shown in the figure) along its path and serves as
the local oscillator (LO) for the homodyne measurements~\cite{reynaud92}. In order to
finely adjust the alignment and the synchronization between the signal and LO pulses,
we use the stimulated beam produced by injecting a different seed pulse into the idler
channel of the parametric crystal. Under appropriate conditions~\cite{lvovsky02:epjd},
the beam generated by stimulated emission on the signal channel is emitted in a spatial
mode which closely matches that of the target signal beam and can thus be used for
alignment purposes. Measurements are performed at different values of the coherent seed
amplitude $|\alpha|$ by rotating the half-wave plate; as seen above, a calibration of
such an amplitude is simply obtained from a measurement of the increase in the idler
count rate.

\section{Time-domain homodyne measurements}

The pulsed homodyne detection scheme used to analyze the quantum states has been
recently developed by our group and is currently the only system capable of operating
at the full repetition rate (80 MHz) of common mode-locked lasers in the time domain
\cite{josab02,pra04}. The fields at the two output ports of the beam-splitter are
detected by two photodiodes (Hamamatsu S3883, with active area 1.7 mm$^2$) whose
difference signal is amplified and sent to a fast digital oscilloscope whose
acquisition is triggered by the detection events in the idler channel. Each acquisition
frame spans two consecutive LO pulses where only the first one is synchronized with the
detection of an idler photon and contains the ``information'' about the SPACS
$\ket{\alpha,1}_s$, while the second one can be used for the measurement of the
reference un-excited coherent state $\ket{\alpha}_s$. By blocking the seed coherent
pulse, the single-photon Fock state $\ket{1}_s$ and the LO shot-noise distributions
corresponding to the vacuum state $\ket{0}_s$ are simultaneously measured. About 5000
acquisition frames can be stored sequentially in the scope at a maximum rate of 160,000
frames per second. Each sequence of frames is then transferred to a personal computer
where the areas of the pulses are measured and their statistic distributions are
analyzed in real time.

If a narrow temporal gate with the laser pulses is used, the typical rate of state preparation
for vacuum input is about 300 s$^{-1}$, with less than 1\% contribution from accidental
counts. A typical sequence of about 5000 acquisition frames can thus be captured and analyzed
in about 20-30 s when no coherent seed is fed in the signal mode. It is interesting to remind
that the probability of detecting an idler photon is proportional to $|\aop^\dag
\ket{\alpha}|^2$, hence, as soon as $\alpha$ is increased and stimulated emission starts
taking place, the rate of trigger events grows proportional to $(1+|\alpha|^2)$, thus making
the acquisition rate much higher. However, even at the maximum values ($|\alpha|\approx 7)$
reached in the experiments, the trigger rate never exceeds $2\cdot10^4 s^{-1}$ (to be compared
with the laser pulse repetition rate of about $8\cdot10^7 s^{-1}$), so that the probability of
conditioning the measurement upon more than a single idler photon always remains negligible.

To explore the different quadratures of the generated field, its phase $\theta$ relative to
the LO field has been varied in controlled steps by applying a voltage to a piezoelectric
transducer (PZT) which slightly translates one of the steering SPACS mirrors. A very good
passive stabilization of the large (about 2 m long) Mach-Zehnder-like interferometer formed by
the paths of the LO and of the seed coherent beam has been achieved in order to guarantee a
constant relative phase during the acquisition of a frame sequence. This is extremely tricky
especially for low values of $\alpha$, where the acquisition rate is lowest. About 50
acquisitions were performed at each value of the coherent seed amplitude $\alpha$ for 10-15
phase values in the $[0,\pi]$ interval.

In order to reduce the contribution from low-frequency noise in the detection system, the
amplified difference signal from the two photodiodes is AC-coupled (cut-off frequency of about
3 kHz at -3 dB) before subsequent amplification and acquisition. This high-pass filter,
combined with the temporal sampling operated by the software for the measurement of the pulse
areas, results in a reduced contribution from the DC terms in the acquired homodyne data which
has to be taken into account in the analysis. While the fast pulse-to-pulse fluctuations which
contribute to the marginal distributions are not affected by such filtering, the mean value of
such distributions has to be scaled by a factor that depends on the sampling window used to
measure the pulse area. This is clearly not an issue when measuring the vacuum field or Fock
states, since the mean value of their marginal distributions is constantly zero, but it has to
be carefully considered when dealing with states having a non-null mean field value. A
calibration of such a factor can be simply carried out by comparing the sequence of marginal
distributions as measured for the coherent state and the value of $\alpha$ as obtained from
the idler counts, and can then be used to re-scale all the mean marginal values before
subsequent analysis. This procedure was double-checked also by fitting the experimental SPACS
marginals to the expected theoretical shapes and obtaining an independent measure of
$|\alpha|$ to compare with the one deduced from the idler counts.

\section{Data analysis and discussion}

Balanced homodyne detection allows the measurement of the signal electric field
quadratures $\hat{x}_\theta=\hat x \cos \theta +\hat y \sin \theta$ as a function of
the relative phase $\theta$ imposed between the LO and the signal, where the two
orthogonal field quadratures $\hat x$ and $\hat y$ are defined as $\hat x =
\frac{1}{2}(\aop +\aop^\dag)$ and $\hat y =\frac{i}{2}(\aop^\dag -\aop)$ and $[\hat
x,\hat y]=i/2$. By performing a series of homodyne measurements on equally-prepared
states it is possible to obtain the probability distributions $p(x,\theta)$ of the
quadrature operator $\hat{x}_\theta=\frac{1}{2}(\aop e^{-i\theta}+\aop^\dag
e^{i\theta})$ that are simply seen to correspond to the marginals of the Wigner
quasi-probability distribution $W(x,y)$~\cite{vogel89}:
\begin{equation}
p(x,\theta) =
\int_{-\infty}^{+\infty}
W(x\cos \theta-y\sin\theta, x\sin \theta+y\cos\theta)\ud y.
\label{eq:radon}
\end{equation}
Given a sufficient number of quadrature distributions at different values of the phase $\theta
\in [0,\pi]$, one is able to reconstruct the quantum state of the field under
study~\cite{leonhardt97}. The elements of the density matrix $\hat\rho$ of the state in the
number-state representation can be obtained by averaging the so called ``pattern functions''
$f_{nm}(x,\theta)$ over the outcomes of the quadrature operator and over the phase $\theta$ as
\begin{equation}
\bra{n}\hat\rho\ket{m}=\frac{1}{\pi}\int_{0}^{\pi}d\theta\int_{-\infty}^{+\infty}dx\,
p(x,\theta)f_{nm}(x,\theta),
\end{equation}
where the pattern functions can be implemented for unit quantum efficiency with stable
numerical algorithms~\cite{dariano97,leonhardt97}. The Wigner function can then be obtained by
means of the following transformation:
\begin{equation}
W(x,y)=\sum_{n,m}^M \rho_{n,m} W_{n,m}(x,y)
\end{equation}
where $W_{n,m}(x,y)$ is the Wigner function of the operator $\ket{n}\bra{m}$. Note that, using
this procedure, the Wigner function of the state is reconstructed from a truncated density
matrix of dimension $M\times M$. This implies a finite resolution in the reconstructed
function which, however, can be adapted to the particular physical situation of interest in
order to avoid loss of information on the state.

\begin{figure}[h]
\includegraphics[width=85mm]{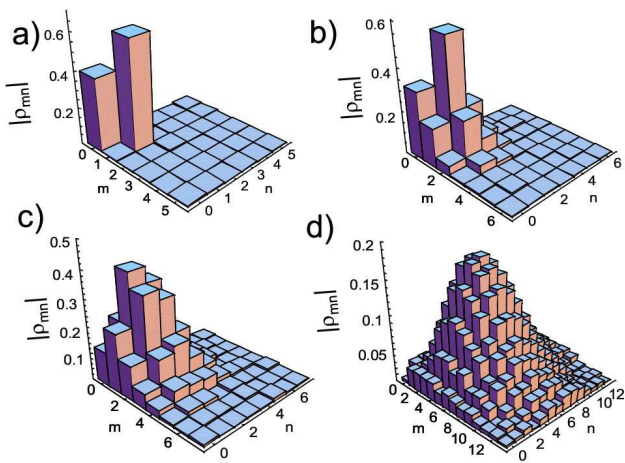}
\caption{(color online) Density matrices of the SPACSs as reconstructed from the experimental
data for increasing values of the seed amplitude. a) $|\alpha|=0$ i.e. single photon Fock
state, b) $|\alpha|=0.387$, c) $|\alpha| = 0.955$, d) $|\alpha| = 2.61$.\label{fig:rec_dm}}
\end{figure}

In Fig.~\ref{fig:rec_dm} a sequence of SPACS reconstructed density matrices is shown for
increasing values of the seed coherent field amplitudes $|\alpha|$.
In Fig.~\ref{fig:rec_wigner} the corresponding Wigner functions, obtained from the truncated
density matrices of Fig~\ref{fig:rec_dm} are shown:
\begin{figure}[h]
\includegraphics[width=85mm]{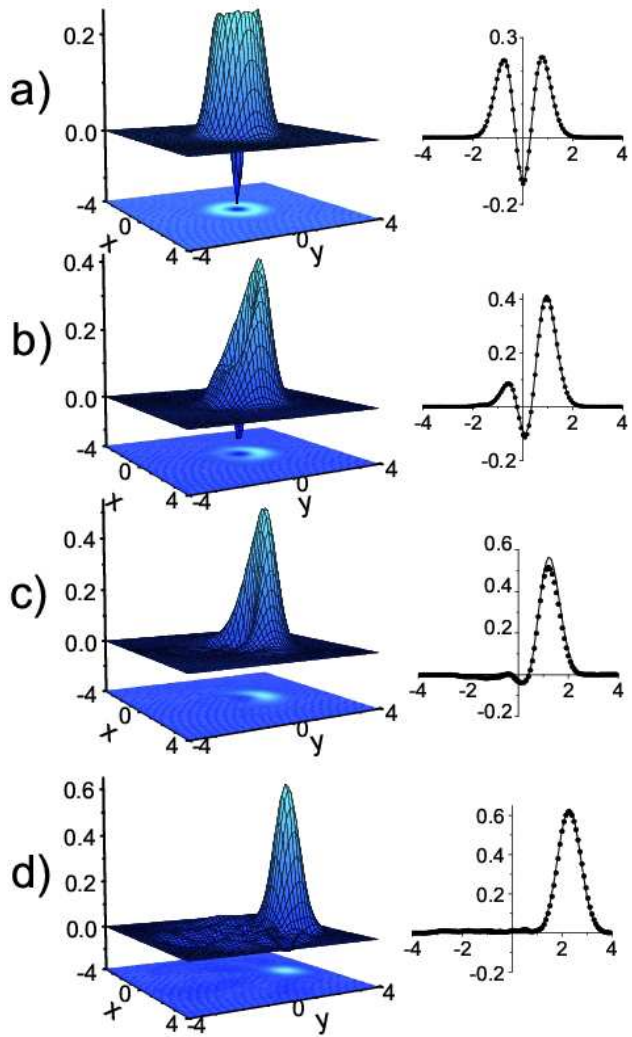}
\caption{(color online) Wigner functions of the SPACSs as reconstructed from the
density matrix elements shown in Fig~\ref{fig:rec_dm}. Also shown are sections of the
reconstructed Wigner functions in the $x=0$ plane (data points), together with the ones
(solid lines) calculated as explained below by taking the limited efficiency of the
system into account.\label{fig:rec_wigner}}
\end{figure}
the first one (a), obtained with a blocked input, corresponds to the single-photon Fock
state obtained by conditional preparation from the two-photon wavefunction of SPDC
\cite{pra04, lvovsky01} and clearly exhibits classically impossible negative values
around the center of the circularly symmetric (due to the undefined value of the phase)
distribution. When the coherent seed is initially switched on at very low intensity
($|\alpha|\approx 0.4$, i.e. an average of one photon every 7 pulses), the Wigner
function starts to loose its circular symmetry while moving away from the origin due to
the gradual appearance of a defined phase, but it still exhibits a clear non-classical
nature as indicated by its partial negativity (b). For increasing seed amplitudes, the
negativity gradually gets less evident (c) and the ring-like wings in the distribution
start to disappear making it more and more similar to the Gaussian typical of a
classical coherent field (d). Interestingly, even at relatively high input amplitude
$\alpha$, the Wigner distribution for the SPACS $\ket{\alpha,1}$ keeps showing the
effect of the one-photon excitation when compared to the corresponding, slightly
displaced, un-excited $\ket{\alpha}$ state~\cite{science04}.

When comparing the reconstructed Wigner functions and density matrix elements to the
theoretical ones for the corresponding quantum states, one has to take into account the
limited efficiency of the homodyne detection apparatus which does not allow one to
generate and analyze pure states but always involves some mixing with the vacuum. The
limited efficiency enters both in the preparation of the quantum state, where the dark
counts and the non-ideal conditioning in the idler channel do not allow one to generate
a completely pure state in the signal channel, both in the homodyne detection process
itself, due to the limited efficiency of the photodiodes and to the imperfect
mode-matching of the signal field with the LO~\cite{pra04}.

Here the limited efficiency can be measured directly by studying the single-photon Fock
state obtained by blocking the seed coherent field. The non-unit efficiency of the
apparatus prevents the observation of the real single-photon Wigner function, and what
one gets instead is its convolution with the vacuum one. The convolution result is the
well known $s$-parametrized quasi-probability distribution with the $s$ parameter
scaled by the detection efficiency $\eta$~\cite{mandelwolf,leonhardt97}. From a fit of
the experimental quadrature distributions to the corresponding theoretical
phase-independent, marginal curves, we obtain an overall efficiency of $\eta = 0.602
\pm 0.002$.

The Wigner function of SPACSs in presence of limited efficiency is:
\begin{equation}
W(z)=\frac{-2[2\eta-1-|2\sqrt{\eta}z-\alpha(2\eta-1)|^2]}{\pi(1+|\alpha|^2)}
e^{-2|z-\sqrt{\eta}\alpha|^2} \label{eq:spacs_wig_eta}
\end{equation}
and one can easily see that, as for single-photon Fock states, negative values can only
be achieved with $\eta>0.5$. The non-unit detection efficiency thus reduces the
non-classical character of experimentally observed SPACSs and, especially for higher
values of the seed amplitude $|\alpha|$, may completely mask it in the presence of
reconstruction noise (see Fig.\ref{fig:rec_wigner}d)). In our case, a good efficiency
combined with relatively low reconstruction errors allow us to clearly observe the
non-classical character of SPACSs up to $|\alpha|\approx 2$.

\begin{figure}[h]
\includegraphics[width=85mm]{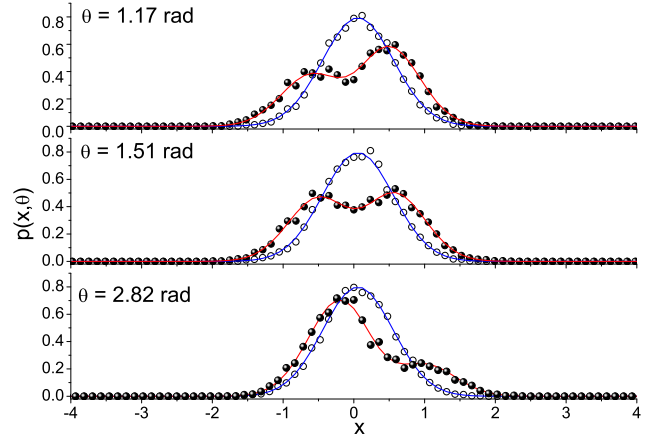}
\caption{(color online) Normalized histograms of the pulse-integrated homodyne signal
for the SPACS (filled circles) and the coherent seed field (empty circles) at different
phases relative to the LO. The quadrature $x$ axis is normalized to the vacuum/coherent
state distribution width. Also plotted (solid curves) are the fits to the theoretical
curves for the Wigner marginals including the effects of the limited efficiency.
\label{fig:marginals}}
\end{figure}

The marginal distributions of the SPACSs in the case of limited efficiency have the
form:
\begin{eqnarray}
\lefteqn{p(x,\theta,\alpha,\eta) =
\frac{1}{1+|\alpha|^2}\sqrt{\frac{2}{\pi}} \Big[ 1-\eta+ 4\eta x^2 {} } \nonumber \\
&& {}+|\alpha|^2(1+2\eta(\eta-1))-4|\alpha| x\sqrt{\eta}(2\eta-1)\cos(\theta) \nonumber \\
&& {}+2|\alpha|^2\eta(\eta-1)\cos(2\theta)\Big] e^{-2(x-|\alpha|\sqrt{\eta}\cos\theta)^2}
\label{eq:spacs_marg_eta}
\end{eqnarray}
and the corresponding mean values and variances have to be modified with respect to the ideal
cases of Eqs.(\ref{eq:spacs_mean_teo}) and (\ref{eq:spacs_var_teo}) as:
\begin{equation}
\mean{x_\theta}_{\alpha,\eta}=
\frac{|\alpha|(2+|\alpha|^2)\sqrt{\eta}\cos(\theta)}{1+|\alpha|^2}
\label{eq:spacs_mean_eta}
\end{equation}
and
\begin{equation}
[\Delta x_\theta]^2_{\alpha,\eta}=
\frac{1}{4}+ \frac{\eta[1-|\alpha|^2\cos{(2\theta)}]}{2(1+|\alpha|^2)^2}
\label{eq:spacs_var_eta}
\end{equation}
Figure~\ref{fig:marginals} presents the experimental marginal distributions of the SPACS at a
fixed value of the coherent seed amplitude $|\alpha|=0.387$ and for three different values of
the phase $\theta$. Superposed to the data points are the curves obtained from a fit of the
distributions to the expected shapes as given by eq.(\ref{eq:spacs_marg_eta}).

\begin{figure}[h]
\includegraphics[width=85mm]{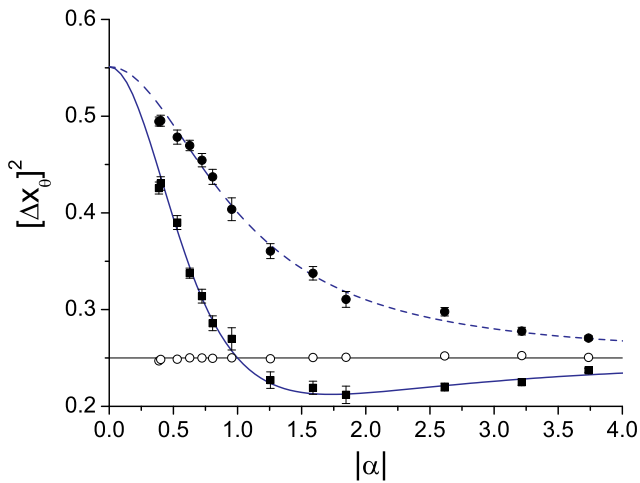}
\caption{(color online) Variances of the squeezed (filled squares) and anti-squeezed
(filled circles) quadratures of the SPACS for different coherent state amplitudes.
Solid and dashed lines are obtained from Eq.~(\ref{eq:spacs_var_eta}) with $\theta=0$
and $\theta=\pi/2$, respectively, and with a global efficiency set to $\eta = 0.6$.
Also shown are the experimental data (empty circles) and the theoretical curve
(horizontal line at 1/4) for the variance of the coherent state. \label{fig:squeezing}}
\end{figure}

The non-classical character of SPACSs is also evident if the quadrature variances are
measured for different amplitudes of the coherent seed field. Indeed, while the
original coherent state has equal fluctuations in the different quadratures
independently from its amplitude, the one-photon-excited state exhibits a squeezing in
one of the quadratures and larger fluctuations in the orthogonal one as soon as
$|\alpha|>1$, as indicated in eqs.(\ref{eq:spacs_var_teo}) and
(\ref{eq:spacs_var_eta}). An intuitive interpretation of this behavior can be connected
with the reduction in the intensity noise of the coherent state when excited by a
perfectly defined number of quanta with the corresponding increase in the phase noise
due to the intrinsic lack of phase information of the Fock state. This effect starts to
become evident in the reconstructed Wigner function of Fig.~\ref{fig:rec_wigner}(c)
(which is however still at the border of the un-squeezed region), where a somewhat
reduced width appears along the radial direction, while the increase in the phase noise
is indicated by the appearance of the ring-like wings in the tangential direction of
the Wigner distribution. A more quantitative measurement of the variance in the
squeezed and anti-squeezed quadratures is presented in Fig.~\ref{fig:squeezing} where
the expected curves for $\theta=0$ and $\theta=\pi/2$ are also drawn according to
eq.(\ref{eq:spacs_var_eta}) with a global efficiency of $\eta = 0.6$.
The experimental variances for the $x_{(\theta=0)}$ quadrature clearly get smaller than
those of the corresponding coherent state (also shown in the graph and independent of
the seed intensity) as soon as the amplitude exceeds unity, and a maximum squeezing of
about 15$\%$ is obtained for $|\alpha|= 1.85$.

\begin{figure}[h]
\includegraphics[width=85mm]{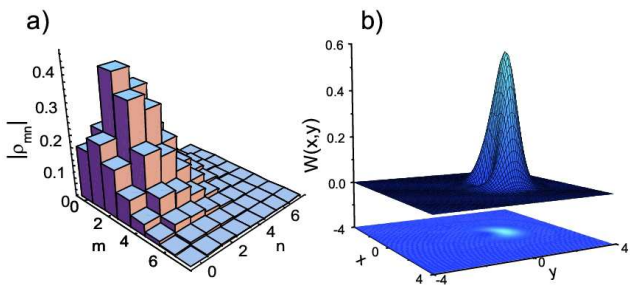}
\caption{(color online) a) Calculated density matrix elements and b) Wigner function of
the SPACS in the case of limited efficiency ($\eta=0.6$) and for $|\alpha|=0.955$.
\label{fig:wigdmcomp}}
\end{figure}

The density matrix elements reconstructed from the data can also be compared with the
theoretical ones provided that a Bernoulli transformation
\begin{equation}
\rho_{i,j}'=\eta^{\frac{i+j}{2}}\sum_{k=0}^{\infty}
\bigg[\binom{i+k}{i}\binom{j+k}{j}\bigg]^{1/2}(1-\eta)^k\rho_{i,j}
\label{eq:spacs_rho_eta}
\end{equation}
is performed in order to include the effects of non-unit efficiency. The expected
density matrix $\rho_c$ is thus obtained from Eqs.(\ref{eq:spacs_rho}) and
(\ref{eq:spacs_rho_eta}) with $\eta=0.6$. Figure~\ref{fig:wigdmcomp} shows the
calculated density matrix elements and the corresponding Wigner function for the SPACS
with $|\alpha|=0.955$.
Such data should be compared to the experimental plots of Figs.~\ref{fig:rec_dm}c) and
~\ref{fig:rec_wigner}c).

\begin{figure}[h]
\includegraphics[width=85mm]{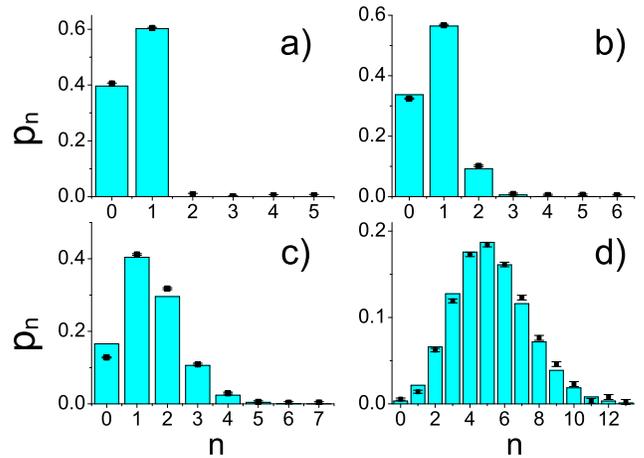}
\caption{(color online) Experimental (data points with error bars) and theoretical
(solid bars) photon number distributions for the SPACS in the case of limited
efficiency ($\eta=0.6$). a) $|\alpha|=0$, b) $|\alpha|=0.387$, c) $|\alpha| = 0.955$,
d) $|\alpha| = 2.61$. \label{fig:dmcomp}}
\end{figure}

The theoretical and experimental photon number distributions for the SPACS with
increasing seed amplitudes $|\alpha|$ are also plotted in Fig.~\ref{fig:dmcomp} where
the errors have been calculated as in Ref.\cite{dariano97}.

In order to better quantify the agreement of the experimental results with the
theoretical ones, we calculated the purity $P$ of the reconstructed states and compared
it with the expected one $P_c$. The purity tends to increase with the amplitude of the
seed coherent field as the final state evolves from a mixture of vacuum and single
photon towards a pure coherent one. Results are presented in Table~\ref{tab:fidelity},
where we also show the fidelity of the reconstructed states to the theoretical ones,
calculated as proposed by Jozsa~\cite{jozsa94} for the comparison between mixed quantum
states
\begin{equation}
F\equiv \left|\mathrm{Tr}\left[\sqrt{\sqrt{\rho_c} \rho_e
\sqrt{\rho_c}}\right]\right|^2 \label{eq:fidelity}
\end{equation}
where $\rho_e$ is the experimentally derived density matrix.
\begin{table}[h]
 \centering \caption{Quantum state indicators for SPACSs. $M$ dimension of the reconstructed density matrix; $P_c$ expected purity for the state; $P$ purity of the reconstructed state; $F$ fidelity between the experimentally reconstructed and the expected state.}
 \begin{ruledtabular}
\begin{tabular}{c c c c c}
  $|\alpha|$ & $M$ & $P_c \equiv \mathrm{Tr}(\rho_c ^2)$ & $P \equiv \mathrm{Tr}(\rho_e ^2)$ & $F$ \\ \hline \\
  0 & 6 & 0.52 & 0.53$\pm$0.01 & 1.009$\pm$0.004 \\
  0.387 & 7 & 0.64 & 0.63$\pm$0.01 &0.990$\pm$0.004 \\
  0.955 & 8 & 0.87 & 0.83$\pm$0.02 &0.976$\pm$0.004 \\
  2.61 & 14 & 0.99 & 1.02$\pm$0.05 &0.995$\pm$0.005 \\
\end{tabular}
\end{ruledtabular}
 \label{tab:fidelity}
\end{table}

A comparison of the fidelity values with the photon number distributions of
Fig.~\ref{fig:dmcomp} seems to indicate that fidelity, quickly saturating to unity even
for somewhat different distributions, is not a very sensitive benchmark for the
closeness of the experimental to the expected states in this case. A similar conclusion
has been recently obtained by the group of Kwiat~\cite{kwiat04} in the case of
depolarized entangled mixed states in the discrete variable domain.

\section{Conclusions}
We have successfully used a conditional preparation technique to generate a new class
of light states whose degree of non-classicality can be continuously tuned between the
extreme situations of pure quantum states and classical ones. Such single-photon-added
coherent states are particularly interesting from a fundamental point of view as they
represent the result of the most elementary excitation of a classical light field. In
this regard, the demonstrated possibility to follow their evolution so closely will
certainly push the experimental research towards the investigation of other interesting
and equally fundamental quantum processes. Single-photon-added coherent states are also
noteworthy because, for the first time to our knowledge, both the typical properties of
a quantum character, i.e. the negativity of the Wigner function combined with a
quadrature squeezing, are simultaneously and clearly observed in a light state. While
the non-classicality criterion based on the squeezing can be in principle fulfilled
without bounds on the detection efficiency, as shown in eq.(\ref{eq:spacs_var_eta}),
the negativity of the Wigner function is a much stricter criterion since it requires
efficiencies higher than 50\%. We have demonstrated the possibility of using our
recently developed technique for high-frequency time-resolved balanced homodyne
detection to reconstruct the density matrix elements and the Wigner functions of the
generated states with an overall efficiency of 60\%. Thanks to the very high
acquisition rates achievable by our system, other non-classical states, particularly
those involving higher number of photons and normally characterized by lower generation
efficiencies, are within reach for a complete tomographic analysis.

\section{Acknowledgments}
This work has been performed in the frame of the ``Spettroscopia laser e ottica quantistica''
project of the Department of Physics of the University of Florence and partially supported by
the Italian Ministry of University and Scientific Research (MIUR), under the FIRB contract
RBNE01KZ94.

\bibliography{Fock_bib}% Produces the bibliography via BibTeX.

\end{document}